\documentclass[preprint,5p,number,times,sort, compress]{elsarticle}

\usepackage{lineno}
\usepackage[colorlinks,
            linkcolor=blue, 
            anchorcolor=blue,
            citecolor=blue, 
            ]{hyperref}

\usepackage{color}
\usepackage{amsmath}
\modulolinenumbers[1]

\usepackage[utf8]{inputenc}
\usepackage{graphicx}
\usepackage{subcaption}
\usepackage{amsmath}
\usepackage{booktabs}

\usepackage{amssymb}
\usepackage{ulem}
\usepackage{multirow}
\usepackage{soul}
\setstcolor{red}

\usepackage{makecell}

\journal{NST}

\begin{document}
\begin{frontmatter}

\title{Refined Global Fit of KamLAND Data and the Daya Bay Antineutrino Energy Spectrum}
\author{Guihong Huang\corref{mail}}
\ead{huanggh@wyu.edu.cn}
\address{Wuyi University, Jiangmen 529020, China}

\begin{abstract}
Recently, the JUNO experiment published its measurement of the solar neutrino oscillation parameters $\Delta m^2_{21}$ and $\sin^2\theta_{12}$ based on 59 days of data, with central values differing by $0.2\sigma$ from those released by the KamLAND experiment in 2013. Meanwhile, short-baseline reactor neutrino oscillation experiments such as Daya Bay, RENO, and Double Chooz have observed significant deviations between the measured antineutrino spectrum and the Huber-M\"{u}ller model around 5~MeV. To further investigate the impact of these deviations on the measurement of neutrino oscillation parameters, a global analysis framework is constructed that is weakly dependent on the reactor antineutrino flux model. This framework is based on the independently measured $^{235}\mathrm{U}$ and $^{239}\mathrm{Pu}$ fission antineutrino spectra from the Daya Bay experiment, combined with the public data from KamLAND. First, using the Huber-M\"{u}ller model, the KamLAND 2013 results are successfully reproduced to within $0.1\sigma$. Then, replacing the Huber-M\"{u}ller model with the Daya Bay measured antineutrino spectra in a combined analysis, it is found that the best-fit value of the mass-squared difference $\Delta m^2_{21}$ decreases from $7.53^{+0.17}_{-0.16}\times10^{-5}\,\mathrm{eV^2}$ to $7.50^{+0.19}_{-0.18}\times10^{-5}\,\mathrm{eV^2}$, while the best-fit value of the mixing angle $\tan^2\theta_{12}$ also shows a decreasing trend. This result is in better agreement with the latest JUNO measurement, suggesting that differences in the predicted reactor antineutrino spectra may be an important cause of the tension between the two experiments.
\end{abstract}


\end{frontmatter}

\section{Introduction}
The great success of neutrino oscillation experiments and theories in explaining the solar and atmospheric neutrino anomalies provides strong evidence that neutrinos have non-zero mass. The solar neutrino experiment SNO measured the mass-squared difference $\Delta m^2_{21} = 5.6^{+1.9}_{-1.4}\times10^{-5}\,\mathrm{eV^2}$ and mixing angle $\tan^2\theta_{12} = 0.427^{+0.030}_{-0.029}$ \cite{SNO:2011}. In contrast, the long-baseline reactor neutrino oscillation experiment KamLAND reported $\Delta m^2_{21}=7.54^{+0.19}_{-0.18}\times10^{-5}\,\mathrm{eV^2}$ and $\tan^2\theta_{12}=0.481^{+0.092}_{-0.080}$ \cite{KamLAND2013}. With the constraint on $\theta_{13}$, the KamLAND results become $\Delta m^2_{21}=7.54^{+0.18}_{-0.18}\times10^{-5}\,\mathrm{eV^2}$ and $\tan^2\theta_{12}=0.460^{+0.063}_{-0.056}$ (from Fig.~4 in \cite{KamLAND2013}). Hence, there is some tension between KamLAND and SNO in both $\Delta m^2_{21}$ and $\tan^2\theta_{12}$, especially for $\Delta m^2_{21}$. Very recently, the JUNO experiment published the most precise measurement of solar neutrino oscillation parameters to date, with $\Delta m^2_{21}=7.50^{+0.12}_{-0.12}\times10^{-5}\,\mathrm{eV^2}$ and $\tan^2\theta_{12}=0.4476^{+0.0182}_{-0.0182}$ \cite{JUNO:2025}. This differs from the KamLAND result by $0.2\sigma$, indicating a mild tension.

\indent On the other hand, short-baseline reactor neutrino oscillation experiments have revealed differences between the measured reactor antineutrino spectrum and the Huber-M\"{u}ller model prediction~\cite{Mueller, Huber}, both in shape and absolute flux. These differences manifest as an approximately $6\%$ deficit in the total reactor neutrino flux and the so-called ``5~MeV bump'' \cite{DC:2014, RENO:2015, DayaBay:2017, DayaBay:2019, dayabay:2021}. Given the central role of the reactor antineutrino spectrum in reactor neutrino oscillation analyses, it is of great interest to understand how the spectral anomaly affects the fitted neutrino mass and mixing parameters. A global analysis reported $\Delta m^2_{21}=7.34^{+0.17}_{-0.16}\times10^{-5}\,\mathrm{eV^2}$ and $\tan^2\theta_{12}=0.437^{+0.029}_{-0.026}$ \cite{nufit:2016}, where incorporating the bump feature lowered $\Delta m^2_{21}$ by about $1\sigma$. That study used the overall energy-dependent ratio $f(E)$ between data\cite{DayaBay:2017} and the Huber-M\"{u}ller expectation, and subsequent work adopted the Daya Bay unfolded antineutrino spectrum \cite{dayabay:2021}, but without a quantitative discussion of its impact on solar neutrino parameters \cite{nufit:2024}. To more accurately assess how the reactor neutrino flux expectation influences the KamLAND fit, we construct in this paper a global analysis framework that is weakly dependent on the reactor neutrino flux model.

\indent  In this study, the latest KamLAND analysis results are first reproduced. Then, by treating the $^{235}\mathrm{U}$ and $^{239}\mathrm{Pu}$ antineutrino spectra as free parameters and imposing constraints from the Daya Bay experiment, a refined combined fit of KamLAND data and Daya Bay antineutrino spectra is performed. For completeness, a simplified spectral correction method similar to that in Ref.~\cite{nufit:2016} is also investigated. The final results show that in the combined fit, $\Delta m^2_{21}$ decreases by about $0.17\sigma$, which is slightly smaller than the $0.29\sigma$ decrease obtained with the simplified correction. The global analysis framework developed here will be valuable for understanding the mild tension between JUNO and KamLAND, as well as for future applications of reactor antineutrino spectra, such as the unfolded spectrum from JUNO-TAO~\cite{JUNO-TAO, JUNONMO}.

\section{Reproducing the KamLAND Results}
The expected number of reactor neutrino events is determined by five key ingredients: the expected antineutrino spectrum $\Phi_{j}(E_{\nu})$, the antineutrino survival probability $P_{j}(E_{\nu},\Delta m^2_{21}, 
\theta_{12},  \theta_{13})$, the inverse beta decay (IBD) cross section $\sigma(E_{\nu})$ \cite{CS}, the detector response $R(E_{e}, E_{e}^{'}, \epsilon^{E})$, and a normalization factor $f$. The expected events in each energy bin is given by:
\begin{align}
    &N^{th}_{i}(\Delta m^2_{21}, \theta_{12}, \theta_{13}) =f\int \mathrm{d}E_{\nu}\,\sigma(E_{\nu})\sum_{j} \Phi_{j}(E_{\nu})\times \nonumber\\
    &P_{j}(E_{\nu},\Delta m^2_{21}, \theta_{12}, \theta_{13})\int_{i} \mathrm{d}E_{e}\,R(E_{e}, E_{e}^{'}, \epsilon^{E}).\nonumber\\
    &\Phi_{j}(E_{\nu}) =\frac{1}{4\pi L^{2}_{j}} \sum_{k} \frac{f_{k}\cdot W^{th}_{j}}{\sum_{l}f_{l}\cdot e_{l}} \frac{\mathrm{d}\phi_{k}(E_{\nu})}{\mathrm{d}E_{\nu}}.\nonumber\\
    &R(E_{e}, E_{e}^{'}, \epsilon^{E}) = \frac{1}{\sqrt{2\pi}\sigma_{E_{e}^{'}}(1+\epsilon^{E})} \exp\left( -\frac{\left( E_{e} - E_{e}^{'}(1+\epsilon^{E}) \right)^2}{2\sigma^2_{E_{e}^{'}}(1+\epsilon^{E})}\right).\nonumber\\
    &P_{j}(E_{\nu},\Delta m^2_{21}, \theta_{12},  \theta_{13})  = \cos^4\theta_{13} \left( 1-\sin^2(2\theta_{12M})\sin^2\frac{\Delta m^2_{21M}L_j}{4E_{\nu}} \right) \nonumber\\
    &+\sin^4\theta_{13}.
\end{align}
Here $d\phi_{i}(E_{\nu})/dE_{\nu}$ is the antineutrino spectrum, $W^{th}_{j}$ the thermal power of reactor $j$, $L_{j}$ the baseline length, and $f_{k}$ the fission fraction of isotope $k$ in reactor $j$. The Huber-M\"{u}ller model is used to compute $d\phi_{k}(E_{\nu})/dE_{\nu}$. The detector response is a function of the prompt signal energy $E_{e}$, the true positron energy $E_{e}^{'}$, and the linear energy scale offset $\epsilon^{E}$. The quantities $\theta_{12M}$ and $m^2_{21M}$ are the effective mixing angle and mass-squared difference in matter, incorporating the Mikheyev–Smirnov–Wolfenstein (MSW) effect~\cite{Wolfenstein, Mikheyev}.

\begin{figure}[t]
\centering
\includegraphics[width=0.9\linewidth]{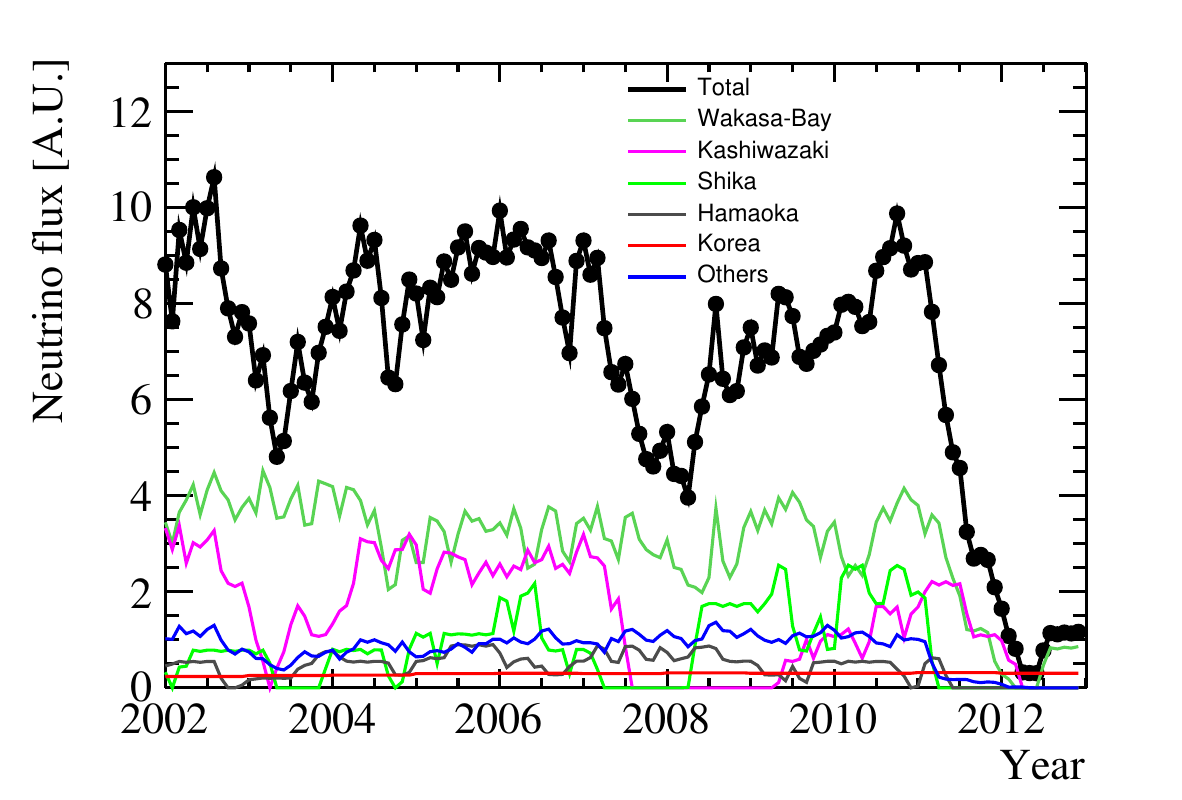}
\caption{Reactor antineutrino flux at the KamLAND site calculated from the electricity generation data of Japanese and South Korean nuclear power plants. The total flux and contributions from different reactors are shown, in good agreement with the KamLAND results \cite{KamLAND2013}.}
\label{fig:nvflux}
\end{figure}

The key to computing the expected event rate is the neutrino flux. Two websites \cite{Flux} record the monthly electricity generation of Japanese nuclear power plants and the annual generation of South Korean plants. Since the electricity output is approximately proportional to the thermal power of a reactor, these data are used to estimate the relative neutrino flux from the 22 reactors in Japan and South Korea. As an important cross-check, Fig.~\ref{fig:nvflux} shows our calculated flux, which agrees well with the KamLAND result \cite{KamLAND2013}.

Next, we acquire the experimental data. In the KamLAND 2013 analysis, data were divided into three running periods for a combined analysis \cite{KamLAND2013}. Without neutrino oscillation, the expected number of reactor antineutrino events is 3564. The observed events, experimental errors, best-fit background values, and best-fit geo-neutrino values are given in the reference. The average fission fractions, background uncertainties, and rate systematics are also provided.

\begin{figure}[t]
\centering
\includegraphics[width=0.8\linewidth]{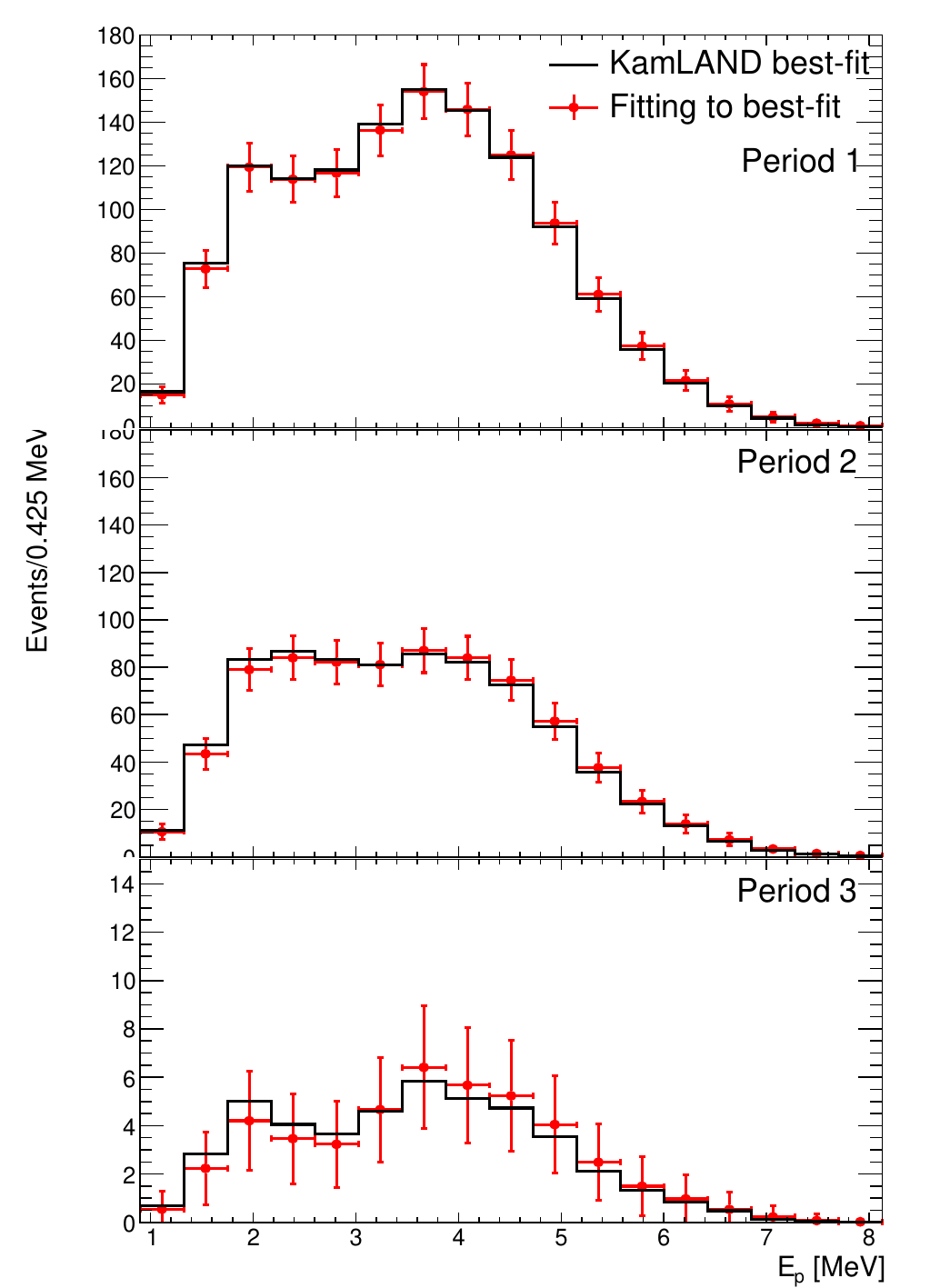}

\caption{Fit results for the best-fit oscillation spectrum. The oscillated spectral shape and the expected no-oscillation events are in good agreement with the official results.}
\label{fig:fit2Best}
\end{figure}

Because the official reference does not provide the expected no-oscillation events for each period, an attempt is made to recover these numbers by performing a concise $\chi^2$ fit to the published best-fit oscillation signal with $\Delta m^2_{21}=7.54\times10^{-5}\,\mathrm{eV^2}$, $\tan^2\theta_{12}=0.481$ and $\sin^2\theta_{13}=0.01$. The $\chi^2$ is constructed as 
\begin{align}
\chi^2 = \sum_{k=1}^{3} \sum_{i=1}^{17} \frac{\left(N_{k,i}^{\text{th}}(f_k) - N_{k,i}^{\text{best}}\right)^2}{N_{k,i}^{\text{th}}}.
\label{Eq:EventCountsFit}
\end{align}
In the equation, \(k\) indexes the three KamLAND data periods (P1, P2, P3). The term \(N_{k,i}^{\text{th}}\) represents the expected events predicted by the model for period \(k\) in energy bin \(i\), while \(N_{k,i}^{\text{best}}\) is the corresponding observed events from the reference. The fit results are shown in Fig.~\ref{fig:fit2Best}. The expected no-oscillation events for the three periods are obtained as 2112.31, 1374.24, and 76.66, respectively. Without imposing period-to-period constraints, the total is very close to the official value. However, the fit for the third period is relatively poor, with an absolute deviation of less than one event but a relative deviation of about $20\%$. Given that the third period accounts for only $2\%$ of the total statistics, this period is discarded to avoid artificially amplifying the bias due to rough handling.

This work constructs a $\chi^2$ function as shown in Eq.~(2) based on a likelihood analysis to fit the solar neutrino oscillation parameters. The observables are the events in 17 equal prompt energy bins from 0.900 to 8.125~MeV.
\begin{align}
    \chi^2 &= \sum^{2}_{k=1} \sum^{17}_{i=1} \left( \frac{N^{\mathrm{obs}}_{i,k} - N^{\mathrm{th}}_{i,k}(1+c^{\mathrm{syst}}_{k}\epsilon^{\mathrm{syst}}+\epsilon^{\mathrm{shape}}_{i}) - N^{\mathrm{B}'}_{i,k} - N^{\mathrm{G}'}_{i,k}}{\sqrt{N^{\mathrm{obs}}_{i,k}}} \right)^2\nonumber\\
        &+ \left(  \frac{\epsilon^{\mathrm{syst}}}{\sigma^{\mathrm{syst}}} \right)^2 + \left( \frac{\epsilon^{E}}{\sigma^{E}} \right)^2
       + \left(  \frac{\epsilon^{\mathrm{Li}}}{\sigma^{\mathrm{Li}}} \right)^2
       + \left(  \frac{\epsilon^{\alpha \mathrm{nA}}}{\sigma^{\alpha \mathrm{nA}}} \right)^2 + \left(  \frac{\epsilon^{\alpha \mathrm{nB}}}{\sigma^{\alpha \mathrm{nB}}} \right)^2\nonumber\\
        &+ \sum^{17}_{i=1,j=1} \epsilon^{\mathrm{shape}}_{i}H^{-1}_{ij}\epsilon^{\mathrm{shape}}_{j},\nonumber\\
N^{\mathrm{B}'}_{i,k} &= N^{\mathrm{Li}}_{i,k}(1+c^{\mathrm{Li}}_{k}\epsilon^{\mathrm{Li}}) + N^{\alpha \mathrm{nA}}_{i,k}(1+c^{\alpha \mathrm{nA}}_{k}\epsilon^{\alpha \mathrm{nA}}) \nonumber\\
&+ N^{\alpha \mathrm{nB}}_{i,k}(1+c^{\alpha \mathrm{nB}}_{k}\epsilon^{\alpha \mathrm{nB}}) + N^{\mathrm{Acc}}_{i,k},\nonumber\\
N^{\mathrm{G}'}_{i,k} &= N^{\mathrm{Geo}}_{i,k}(1+\epsilon^{\mathrm{Geo}}),
\end{align}
where $N^{\mathrm{obs}}_{i,k}$ is the observed events in energy bin $i$ for period $k$, $N^{\mathrm{th}}_{i,k}$ is the expected reactor neutrino events, $N^{\mathrm{B}'}_{i,k}$ is the expected events after correcting for main backgrounds, including $^{9}\text{Li}/^{8}\text{He}$, $^{13}\mathrm{C}(\alpha,n)^{16}\mathrm{O}_{\mathrm{g.s.}}$ ($\alpha$nA), $^{13}\mathrm{C}(\alpha,n)^{16}\mathrm{O}^{*}$ ($\alpha$nB), and accidental coincidences, and $N^{\mathrm{G}'}_{i,k}$ is the geo-neutrino background contribution. $N^{\mathrm{Li}}_{i,k}$, $N^{\alpha\mathrm{nA}}_{i,k}$, $N^{\alpha\mathrm{nB}}_{i,k}$, and $N^{\mathrm{Acc}}_{i,k}$ are the expected events for each background component. $\sigma^{\mathrm{syst}}$ is the overall systematic uncertainty, $\sigma^{E}$ is the energy scale uncertainty, and $\sigma^{\mathrm{Li}}$, $\sigma^{\alpha\mathrm{nA}}$, $\sigma^{\alpha\mathrm{nB}}$ are the background uncertainties. The corresponding pull parameters are $\epsilon^{\mathrm{syst}}$, $\epsilon^{E}$, $\epsilon^{\mathrm{Li}}$, $\epsilon^{\alpha\mathrm{nA}}$, and $\epsilon^{\alpha\mathrm{nB}}$, respectively. The bin-to-bin spectral shape uncertainties are constrained by the covariance matrix $H_{ij}$ with pull parameters $\epsilon^{\mathrm{shape}}_i$. The values of uncertainties are listed in Table~\ref{tab:uncertainties}. The correlation factors $c^{\text{syst}}_{k}$, $c^{\mathrm{Li}}_k$, $c^{\alpha\mathrm{nA}}_k$, $c^{\alpha\mathrm{nB}}_k$ are defined as $\sigma^{\text{syst}}_k / \sigma^{\text{syst}}_1$, $\sigma^{\mathrm{Li}}_k / \sigma^{\mathrm{Li}}_1$, $\sigma^{\alpha\mathrm{nA}}_k / \sigma^{\alpha\mathrm{nA}}_1$, $\sigma^{\alpha\mathrm{nB}}_k / \sigma^{\alpha\mathrm{nB}}_1$, respectively, reflecting the correlation of the uncertainties between periods. 

The quadratic terms in the equation are penalty terms for systematic error parameters, and $\sum_{i,j} \epsilon^{\text{shape}}_{i} H^{-1}_{ij} \epsilon^{\text{shape}}_{j}$ constrains the shape parameters. Penalty terms are introduced to constrain the energy scale, backgrounds, and bin-to-bin shape variations. Following KamLAND's treatment, we assume that systematic uncertainties are fully correlated between the two periods, and the geo-neutrino background is allowed to vary freely. The KamLAND detector-related uncertainty parameters used in this study are listed in Table~\ref{tab:uncertainties}, all taken from the official publication.

\begin{table}[t]
\centering
\caption{Relevant systematic uncertainties for KamLAND data (in \%)}
\label{tab:uncertainties}
\begin{tabular}{lccccc}
\toprule
& \multicolumn{5}{c}{Uncertainty source} \\
\cmidrule(lr){2-6}
Phase & $\sigma^{\mathrm{syst}}$ & $\sigma^{E}$ & $\sigma^{\mathrm{Li}}$ & $\sigma^{\mathrm{\alpha nA}}$ & $\sigma^{\mathrm{\alpha nB}}$ \\
\midrule
P1 & 3.0 & 2.0 & 8.0 & 10.0 & 20.0 \\
P2 & 3.5 & 2.0 & 10.0 & 23.0 & 30.0 \\
P3 & 3.5 & 2.0 & 20.0 & 43.0 & 45.0 \\
\bottomrule
\end{tabular}
\end{table}

\begin{figure}[t]
\centering
\includegraphics[width=0.8\linewidth]{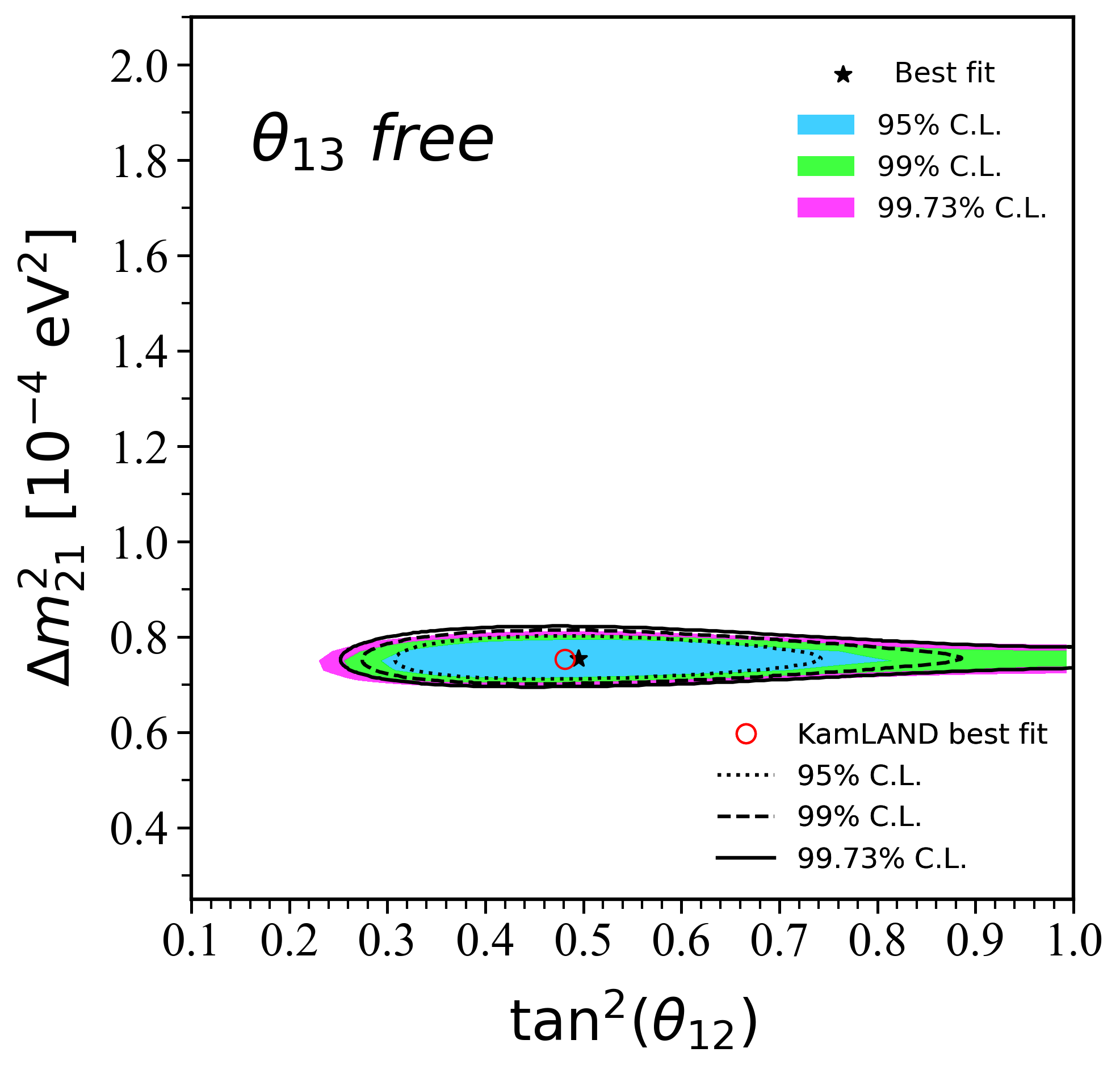}
\includegraphics[width=0.8\linewidth]{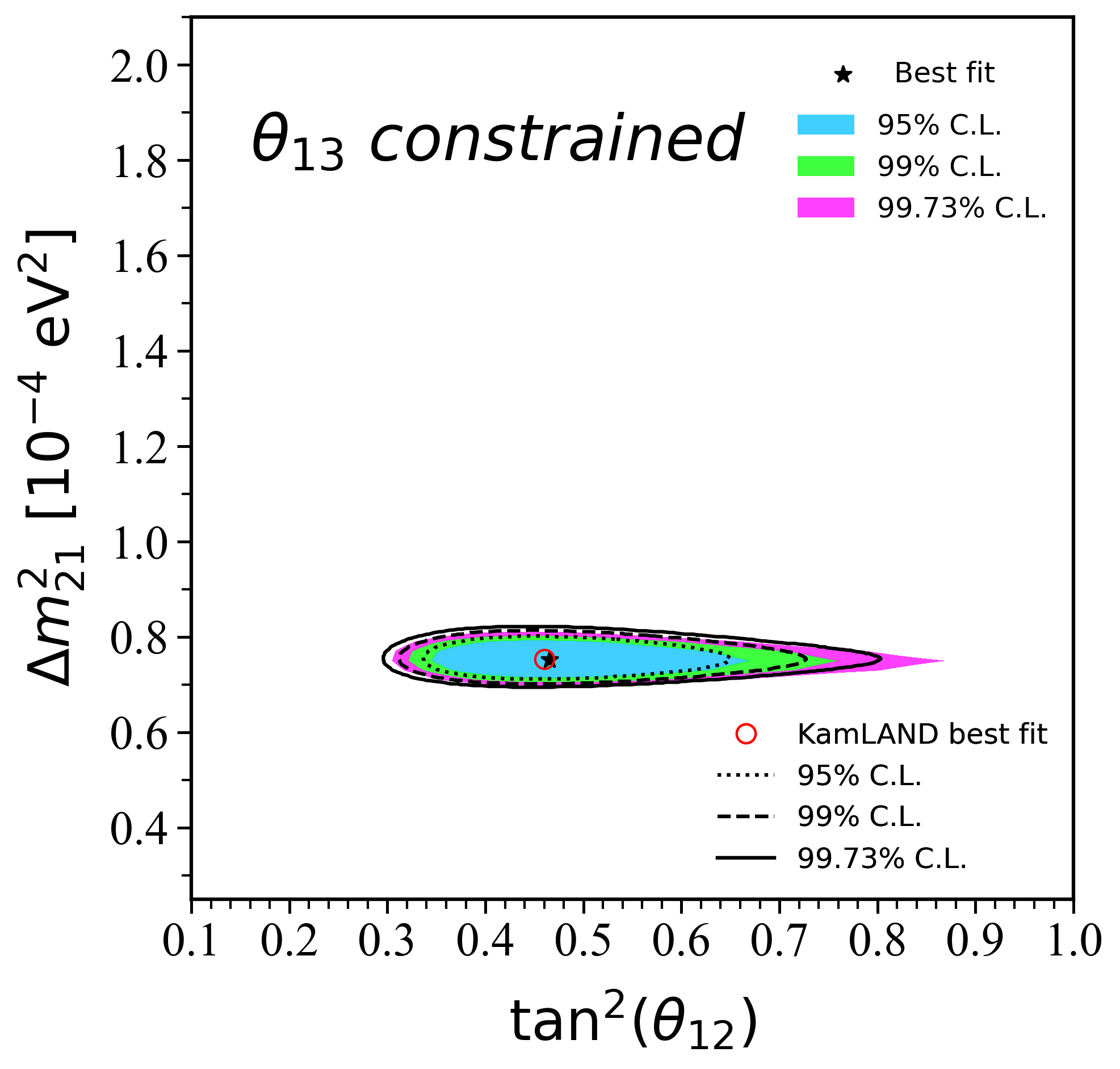}
\caption{Allowed regions at $95\%$, $99\%$, and $99.73\%$ confidence levels in the ($\tan^2\theta_{12},\,\Delta m_{21}^2$) plane. The filled colored areas are the reproduction results of this work, and the contours are the official results.}
\label{fig:reprod}
\end{figure}

The reproduced allowed regions in the $\Delta m^2_{21}$--$\tan^2\theta_{12}$ plane are shown in Fig.~\ref{fig:reprod}, and the best-fit values are given in Tab.~\ref{tab:results1}. The official KamLAND results are also shown for comparison. With $\theta_{13}$ free, the reproduced $\Delta m^2_{21}$ and $\sin^2\theta_{13}$ agree with the official results within $0.15\sigma$, while $\tan^2\theta_{12}$ is slightly larger. When $\sin^2\theta_{13}$ is constrained to $0.023^{+0.002}_{-0.002}$, the reproduction becomes even more accurate, agreeing with the official results within $0.1\sigma$. This lays a solid foundation for the subsequent combined analysis with the Daya Bay antineutrino spectra.

\begin{table}[t]
\centering
\caption{Best-fit solar neutrino oscillation parameters from KamLAND official results and this reproduction, with $\theta_{13}$ free or constrained.}\label{Table 1_transposed}
\renewcommand{\arraystretch}{1.5}
\begin{tabular}{>{\centering\arraybackslash}p{2.2cm} c c c}
\toprule
& $\Delta m^2_{21}$ & $\tan^2 \theta_{12}$ & $\sin^2 \theta_{13}$ \\
& [$10^{-5}\,\mathrm{eV^2}$] & & \\
\midrule
$\theta_{13}$ free & & & \\
Reproduced & $7.54^{+0.17}_{-0.17}$ & $0.495^{+0.106}_{-0.095}$ & $0.011^{+0.033}_{-0.035}$ \\
\midrule
$\theta_{13}$ free & & & \\
Official & $7.54^{+0.19}_{-0.18}$ & $0.481^{+0.092}_{-0.080}$ & $0.010^{+0.033}_{-0.034}$ \\
\midrule
$\theta_{13}$ constrained & & & \\
Reproduced & $7.53^{+0.17}_{-0.16}$ & $0.465^{+0.067}_{-0.055}$ & $0.023^{+0.002}_{-0.002}$ \\
\midrule
$\theta_{13}$ constrained & & & \\
Official & $7.53^{+0.18}_{-0.18}$ & $0.460^{+0.063}_{-0.056}$ & $0.023^{+0.002}_{-0.002}$ \\
\bottomrule
\end{tabular}
\label{tab:results1}
\end{table}

\section{Combined Analysis}
\noindent In 2019, the Daya Bay experiment published measurements of the independent antineutrino spectra of $^{235}\mathrm{U}$ and $^{239}\mathrm{Pu}$ \cite{DayaBay:2019}. They found that spectral deviations from theory appear at different confidence levels for the two isotopes. This provides the motivation and input for the combined analysis of KamLAND data and Daya Bay antineutrino spectra.

\indent In this combined analysis, the $^{235}\mathrm{U}$ and $^{239}\mathrm{Pu}$ spectra are treated as free parameters, uniformly binned into 25 neutrino energy bins from 2 to 8~MeV. This introduces 50 new free parameters, which are constrained by the unfolded reactor antineutrino spectra from Daya Bay. The $^{238}\mathrm{U}$ and $^{241}\mathrm{Pu}$ spectra are based on the Huber-M\"{u}ller model, with a bin-to-bin uncertainty of $10\%$ assigned in the KamLAND $\chi^2$ term. The expected no-oscillation events are kept unchanged. The Daya Bay inputs include the unfolded spectra of $^{235}\mathrm{U}$ and $^{239}\mathrm{Pu}$, their covariance matrix \cite{DayaBay:2019}, and the Daya Bay detector model that converts prompt energy to deposited energy \cite{DayaBayE:2018}. The $\chi^2$ function for the combined analysis is constructed as:
\begin{align}
\chi^2 &= \sum^{2}_{k} \sum^{17}_{i} \biggl( 
N^{\mathrm{obs}}_{i,k}
        - N^{\mathrm{th}_{^{235}\mathrm{U},^{239}\mathrm{Pu}}}_{i,k}(1+c^{\mathrm{syst}}_{k}\epsilon^{\mathrm{syst}}) \nonumber\\
        &\quad - N^{\mathrm{th}_{^{238}\mathrm{U},^{241}\mathrm{Pu}}}_{i,k}(1+c^{\mathrm{syst}}_{k}\epsilon^{\mathrm{syst}}+\epsilon^{\mathrm{shape}}_{i})
          - N^{\mathrm{B}'}_{i,k} - N^{\mathrm{G}'}_{i,k} \biggr)^2 \Big/ N^{\mathrm{obs}}_{i,k} \nonumber\\
        &+ \left( \frac{\epsilon^{\mathrm{syst}}}{\sigma^{\mathrm{syst}}} \right)^2
        + \left( \frac{\epsilon^{E}}{\sigma^{E}} \right)^2
        + \left( \frac{\epsilon^{\mathrm{Li}}}{\sigma^{\mathrm{Li}}} \right)^2
        + \left( \frac{\epsilon^{\alpha \mathrm{nA}}}{\sigma^{\alpha \mathrm{nA}}} \right)^2
        + \left( \frac{\epsilon^{\alpha \mathrm{nB}}}{\sigma^{\alpha \mathrm{nB}}} \right)^2 \nonumber\\
        &+ \sum^{17}_{i} \left( \frac{\epsilon^{\mathrm{shape}}_{i}}{\sigma^{\mathrm{shape}}} \right)^2
        + \sum^{52}_{r,s} \left( \frac{N^{\mathrm{th}}_{r}}{N^{\mathrm{extr}}_{r}} - 1 \right)
           B^{-1}_{rs} \left( \frac{N^{\mathrm{th}}_{s}}{N^{\mathrm{extr}}_{s}} - 1 \right).
\end{align}
where:
\begin{itemize}
    \item $N^{\text{th}_{^{235}\text{U},^{239}\text{Pu}}}_{i,k}$: The expected events in energy bin $i$ of period $k$ from $^{235}$U and $^{239}$Pu fission. The spectral shape of this term is constrained by the Daya Bay unfolded reactor antineutrino spectra.
    \item $N^{\text{th}_{^{238}\text{U},^{241}\text{Pu}}}_{i,k}$: The theoretical expected events in energy bin $i$ of period $k$ from $^{238}$U and $^{241}$Pu fission.
    \item $\sigma^{\text{shape}}$: The normalization standard deviation of the spectral shape uncertainty for $^{238}\mathrm{U}$ and $^{241}\mathrm{Pu}$. Following the Huber-M\"{u}ller model, a conservative value of $10\%$ is used.
    \item $N^{\text{th}}_r$: The expected events in the $r$-th prompt energy bin of the Daya Bay experiment. Indices $r$ and $s$ run over the 52 bins of the unfolded antineutrino spectra from Daya Bay (26 bins for each of $^{235}$U and $^{239}$Pu).
    \item $N^{\text{extr}}_r$: The prompt energy spectrum of $^{235}$U and $^{239}$Pu from the Daya Bay experiment.
    \item $B_{rs}$: The covariance matrix of the unfolded $^{235}$U and $^{239}$Pu spectra.
\end{itemize}

The last term in the above expression contains the measured prompt energy spectra of $^{235}\mathrm{U}$ and $^{239}\mathrm{Pu}$ from Daya Bay, which, in the joint fit, constrains the variations of the two isotope spectra and thereby propagates the impact to the oscillation parameters. In this combined analysis, the link between the KamLAND data and the Daya Bay antineutrino spectra is the 50 free spectral parameters.

In addition, a simplified combined analysis is performed that directly uses the unfolded antineutrino spectrum from Daya Bay to correct the KamLAND expectation via the ratio of data to theoretical prediction. This simplification is justified by the fact that the average fission fractions of the two experiments are quite close. For simplicity and conservativeness, the covariance matrix and shape uncertainties of the Huber-M\"{u}ller model are kept unchanged in this simplified analysis.

\begin{figure}[t]
\centering
\includegraphics[width=0.8\linewidth]{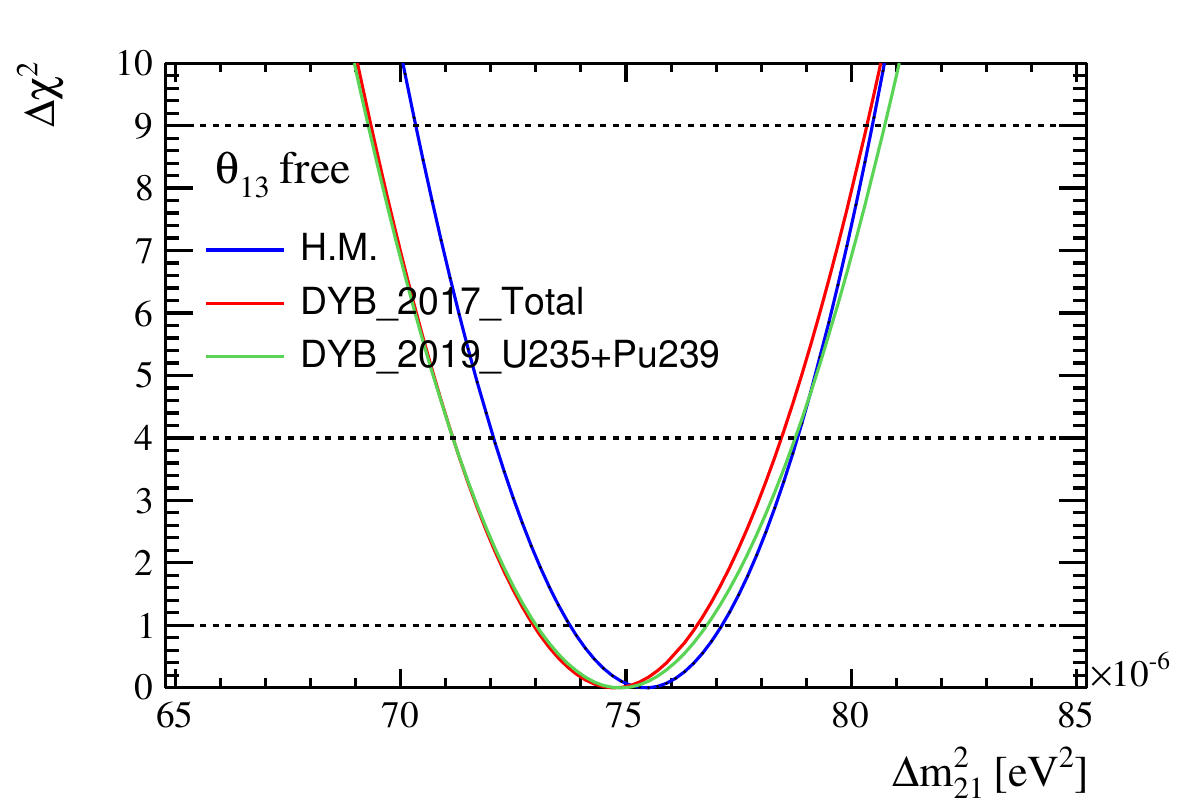}
\includegraphics[width=0.8\linewidth]{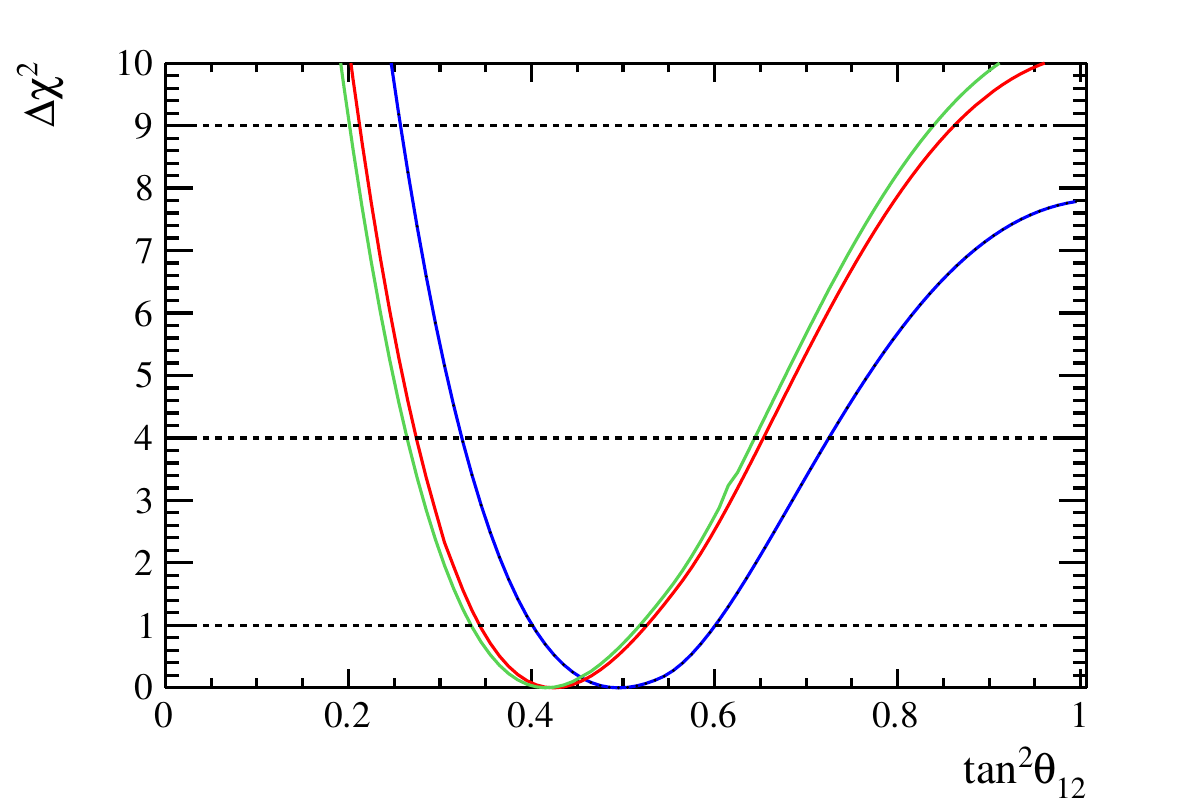}
\includegraphics[width=0.8\linewidth]{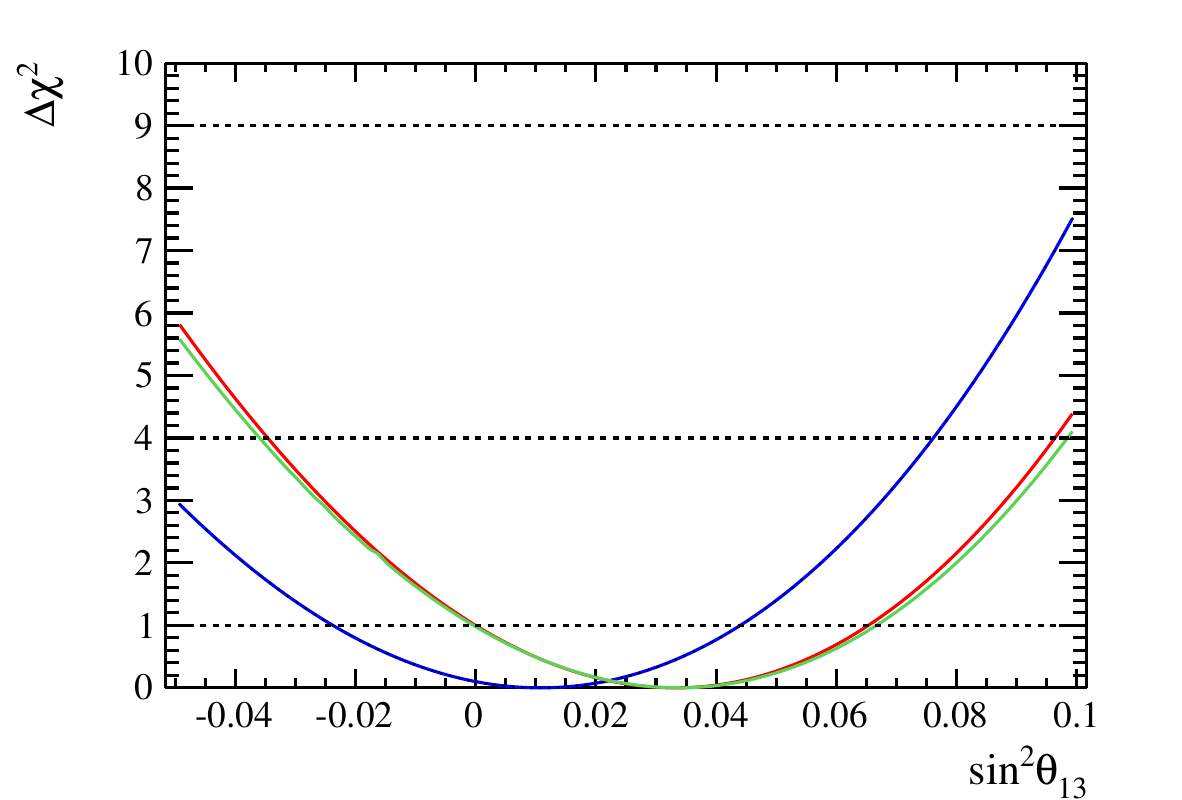}
\caption{Projections of $\Delta \chi^2$ onto each oscillation parameter in the free $\theta_{13}$ case. After including the Daya Bay measured antineutrino spectra, the best-fit central values of $\Delta m^2_{21}$ and $\tan^2\theta_{12}$ decrease systematically.}
\label{fig:Chi2Prof_free13}
\end{figure}

\begin{figure}[t]
\centering
\includegraphics[width=0.8\linewidth]{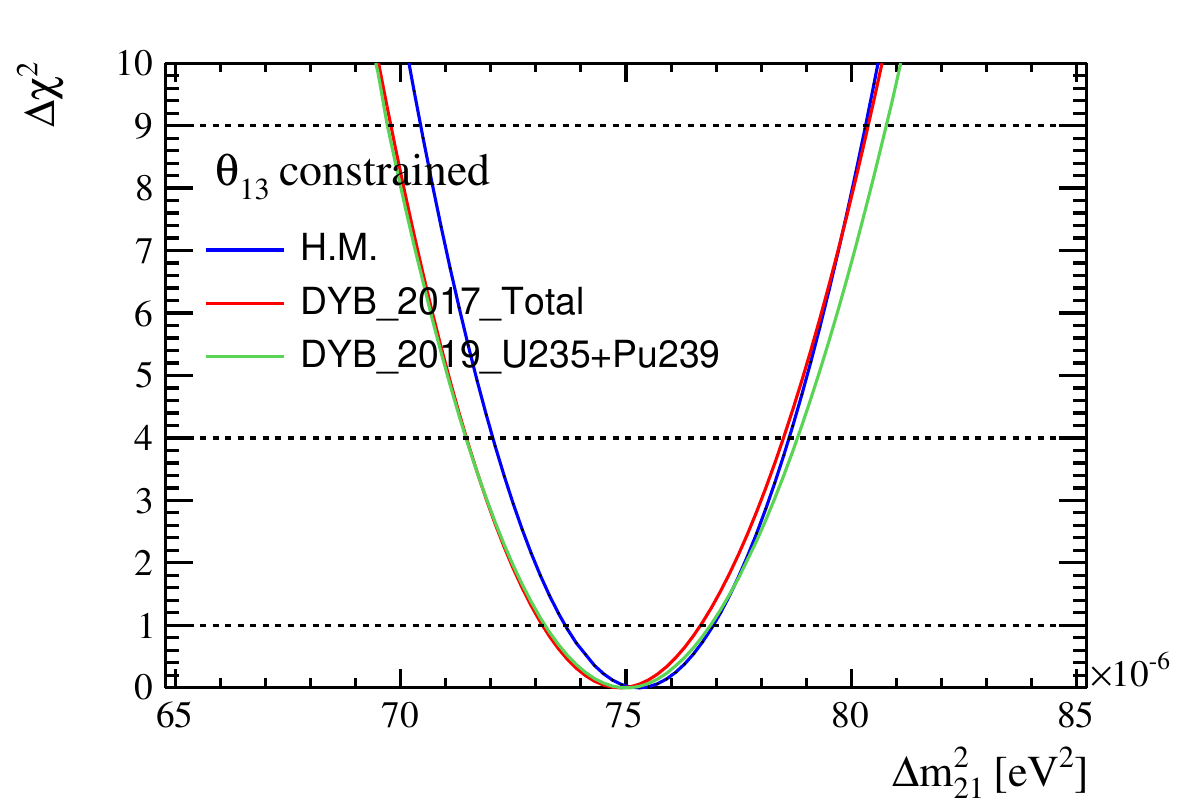}
\includegraphics[width=0.8\linewidth]{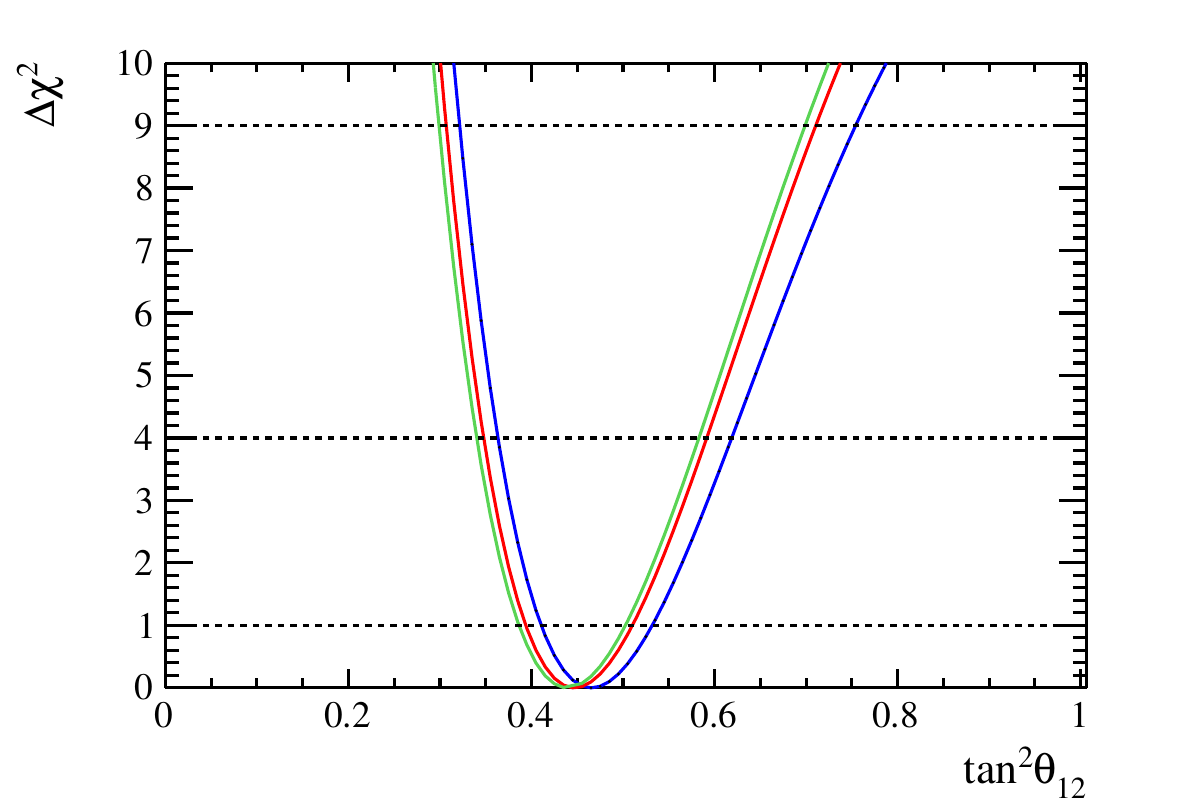}
\caption{Projections of $\Delta \chi^2$ onto each oscillation parameter in the constrained $\theta_{13}$ case. After including the Daya Bay measured antineutrino spectra, the best-fit central values of $\Delta m^2_{21}$ and $\tan^2\theta_{12}$ decrease systematically.}
\label{fig:Chi2Prof_fix13}
\end{figure}

The results of the combined fits are summarized in Figs.~\ref{fig:Chi2Prof_free13},\ref{fig:Chi2Prof_fix13} and Tab.~\ref{tab:results2}. Both combined analysis schemes (with different $\chi^2$ formulations and reactor antineutrino inputs) show a consistent trend in modifying the oscillation parameters: introducing the Daya Bay measured antineutrino spectra systematically lowers the best-fit central values of $\Delta m^2_{21}$ and $\tan^2\theta_{12}$.

The results of the full combined fit using the 2019 Daya Bay independent spectra (+DayaBay19) are first examined. With $\theta_{13}$ free, compared to the pure KamLAND reproduction, $\Delta m^2_{21}$ decreases from $7.54^{+0.17}_{-0.17}\times10^{-5}\,\mathrm{eV^2}$ to $7.49^{+0.19}_{-0.19}\times10^{-5}\,\mathrm{eV^2}$, a shift of approximately $0.26\sigma$ relative to the largest lower uncertainty. $\tan^2\theta_{12}$ drops significantly from $0.495^{+0.106}_{-0.095}$ to $0.416^{+0.103}_{-0.082}$, a shift of about $0.83\sigma$. Meanwhile, $\sin^2\theta_{13}$ increases from $0.011^{+0.033}_{-0.035}$ to $0.034^{+0.033}_{-0.034}$, a change of nearly $1\sigma$. However, given KamLAND's very low sensitivity to $\theta_{13}$, the fitted $\theta_{13}$ has large uncertainties and limited physical significance. In the constrained $\theta_{13}$ mode, $\Delta m^2_{21}$ decreases from $7.53^{+0.17}_{-0.16}\times10^{-5}\,\mathrm{eV^2}$ to $7.50^{+0.19}_{-0.18}\times10^{-5}\,\mathrm{eV^2}$ (a shift of about $0.17\sigma$), and $\tan^2\theta_{12}$ from $0.466^{+0.067}_{-0.055}$ to $0.439^{+0.064}_{-0.053}$ (a shift of about $0.49\sigma$).

Comparing the simplified combined method (+DayaBay17) with the full combined method (+DayaBay19), the trend in oscillation parameters is consistent. In both the free and constrained $\theta_{13}$ cases, the central value of $\Delta m^2_{21}$ from the full fit is slightly higher than that from the simplified fit, with a difference of about $0.01\times10^{-5}\,\mathrm{eV^2}$ ($\sim0.06\sigma$). The central value of $\tan^2\theta_{12}$ from the full fit is slightly lower than that from the simplified fit, with differences of about $0.09\sigma$ and $0.15\sigma$ for free and constrained $\theta_{13}$ cases, respectively. Moreover, the uncertainties in the full fit are slightly larger than those in the simplified fit. This is because the 2019 Daya Bay measurement provides isotope-decomposed spectra and a full covariance matrix, which introduces more independent systematic uncertainties and statistical degrees of freedom compared to the global spectral ratio constraint used in 2017. Consequently, while the central values of the oscillation parameters are suppressed, their uncertainties are moderately enlarged.

\begin{table*}[t]
\centering
\caption{Best-fit results of the oscillation parameters $\Delta m^2_{21}$, $\tan^2\theta_{12}$, and $\sin^2\theta_{13}$ obtained from the KamLAND-only reproduction and from the combined analyses with the Daya Bay 2017 (simplified) and 2019 (full) antineutrino spectra, under both free and constrained $\theta_{13}$ scenarios.}\label{Table 2}
\renewcommand{\arraystretch}{1.5}
\begin{tabular}{ccccccc}
\toprule
& \multicolumn{3}{c}{$\theta_{13}$ free} & \multicolumn{3}{c}{$\theta_{13}$ constrained} \\
\cline{2-7}
    & Reproduction & +DayaBay(17) & +DayaBay(19) & Reproduction & +DayaBay(17) & +DayaBay(19) \\
\hline
    $\Delta m^2_{21}$ [$10^{-5}\,\mathrm{eV^2}$] & $7.54^{+0.17}_{-0.17}$ & $7.48^{+0.18}_{-0.18}$ & $7.49^{+0.19}_{-0.19}$ & $7.53^{+0.17}_{-0.16}$ & $7.49^{+0.18}_{-0.17}$ & $7.50^{+0.19}_{-0.18}$ \\
\hline
    $\tan^2 \theta_{12}$ & $0.495^{+0.106}_{-0.095}$ & $0.425^{+0.102}_{-0.081}$ & $0.416^{+0.103}_{-0.082}$ & $0.466^{+0.067}_{-0.055}$ & $0.447^{+0.064}_{-0.053}$ & $0.439^{+0.064}_{-0.053}$ \\
\hline
    $\sin^2 \theta_{13}$ & $0.011^{+0.033}_{-0.035}$ & $0.033^{+0.032}_{-0.033}$ & $0.034^{+0.033}_{-0.034}$ & $0.023^{+0.002}_{-0.002}$ & $0.023^{+0.002}_{-0.002}$ & $0.023^{+0.002}_{-0.002}$ \\
\bottomrule
\end{tabular}
\label{tab:results2}
\end{table*}

\section{Conclusion}
\noindent Based on the KamLAND 2013 long-baseline reactor neutrino oscillation data and the Daya Bay 2019 short-baseline antineutrino spectra of $^{235}\mathrm{U}$ and $^{239}\mathrm{Pu}$, a detailed combined analysis framework has been constructed that incorporates spectral shape constraints and systematic error correlations. This framework systematically investigates the impact of reactor antineutrino spectrum anomalies on the neutrino oscillation parameters $\Delta m^2_{21}$ and $\tan^2\theta_{12}$. In the constrained $\theta_{13}$ mode, the best-fit solar oscillation parameters are $\Delta m^2_{21}=7.50^{+0.19}_{-0.18}\times 10^{-5}\,\mathrm{eV^2}$ and $\tan^2\theta_{12}=0.439^{+0.064}_{-0.053}$. Compared to the KamLAND standalone fit using the Huber-M\"{u}ller theoretical spectrum, the central values of $\Delta m^2_{21}$ and $\tan^2\theta_{12}$ decrease by about $0.2\sigma$ and $0.4\sigma$, respectively, bringing the measurements into better agreement with the latest high-precision results from JUNO.

This study demonstrates that measurements of the reactor antineutrino spectrum can effectively break the dependence of reactor neutrino oscillation experiments on spectral models and partially alleviate the tension between KamLAND, JUNO and SNO. Furthermore, the combined KamLAND–Daya Bay analysis method developed here can serve as a reference for spectral model corrections in reactor neutrino experiments such as JUNO.

\section*{Acknowledgements}
This work was partially supported by the National Natural Science Foundation of China (Grant No.12405231), the Characteristic Innovation Project for Regular Higher Education Institutions of Guangdong Provincial Department of Education (2024KTSCX044), and the Science Foundation of High-Level Talents of Wuyi University (2021AL027).

\end{document}